\documentclass[10pt]{article}
\usepackage{graphicx}
\usepackage{amsmath}
\usepackage{amssymb}
\usepackage{caption2}
\begin{document}
\bibliographystyle{prsty}
\begin{center}
{\large {\bf \sc{  Analysis of the masses and  decay constants of the heavy-light  mesons with  QCD sum rules }}} \\[2mm]
Zhi-Gang Wang \footnote{E-mail,zgwang@aliyun.com.  }     \\
 Department of Physics, North China Electric Power University,
Baoding 071003, P. R. China
\end{center}

\begin{abstract}
In this article, we calculate the contributions of the vacuum condensates up to dimension-6  including  the $\mathcal{O}(\alpha_s)$ corrections  to the quark condensates in the operator product expansion, then  study the masses and decay constants of the  pseudoscalar, scalar, vector and axial-vector
heavy-light mesons with the QCD sum rules in a systematic way. The masses of the observed  mesons  $(D,D^*)$, $(D_s,D_s^*)$, $(D_0^*(2400),D_1(2430))$, $(D_{s0}^*(2317),D_{s1}(2460))$, $(B,B^*)$, $(B_s,B_s^*)$ can be well reproduced, while the predictions for the masses of  the  $(B^*_{0}, B_{1})$ and  $(B^*_{s0}, B_{s1})$ can be confronted with the experimental data in the future.  We obtain the decay constants of the   pseudoscalar, scalar, vector and axial-vector heavy-light mesons, which  have many phenomenological applications in studying the semi-leptonic and leptonic decays  of the
 heavy-light mesons.
\end{abstract}



\section{Introduction}
The charged heavy-light mesons   can decay to a charged lepton pair $\ell^+ {\nu}_\ell$ through  a virtual
$W^+$ boson. Those leptonic decays  are excellent subjects in studying  the CKM matrix elements and serve as  a powerful probe of new physics beyond the standard model in a complementary way to the direct searches. For example,  the decay widths of the pseudoscalar (P) and vector (V) heavy-light  mesons can be written as
\begin{eqnarray}
\Gamma(P\to \ell\nu) &=& {{G_F^2}\over 8\pi}f_{P}^2  m_{\ell}^2m_{P}\left(1-{m_{\ell}^2\over m_{P}^2}\right)^2 \left|V_{q_1q_2}\right|^2\, , \nonumber\\
\Gamma(V\to \ell\nu) &=& {{G_F^2}\over 12\pi}f_{V}^2  m_{V}^3\left(1-{m_{\ell}^2\over m_{V}^2}\right)^2 \left(1+{m_{\ell}^2\over 2m_{V}^2}\right)\left|V_{q_1q_2}\right|^2\, ,
\end{eqnarray}
in the lowest order approximation, where the $m_{P/V}$ and $f_{P/V}$ are the masses and decay constants, respectively,  the $m_{\ell}$ is the $\ell$
mass, the $V_{q_1 q_2}$ is the CKM  matrix element between the constituent quarks $q_1\bar{q}_2$, and the $G_F$ is the Fermi coupling constant.
If we take the CKM matrix element $V_{q_1 q_2}$ and the branching fractions of the leptonic decays from the CLEO, BaBar, Belle collaborations  as input parameters, then the average values
  $f_D=(204.6 \pm 5.0)\,\rm{MeV}$, $f_{D_s}=(257.5 \pm 4.6)\,\rm{MeV}$ and $f_{D_s}/f_D=1.258 \pm 0.038$ are obtained \cite{PDG}. It is difficult to reproduce the three values consistently in theoretical calculations, such as the QCD sum rules \cite{QCDSRfD2,QCDSRfD5,QCDSRfD4,QCDSRfD6}  and lattice QCD \cite{LattfD1,LattfD2,LattfD3}.  The  discrepancies between the theoretical values   and experimental data  maybe signal some new physics beyond the standard model \cite{fD-Narison-2008}.
 In Ref.\cite{WangJHEP}, we observe that if we take into account the $\mathcal{O}(\alpha_s^2)$ corrections to the perturbative terms and the $\mathcal{O}(\alpha_s)$ corrections to the quark condensate terms and choose the pole masses, the predictions $f_D=(211 \pm 14)\,\rm{MeV}$, $f_{D_s}=(258 \pm 13)\,\rm{MeV}$ and $f_{D_s}/f_D=1.22 \pm 0.08$  are in excellent agreement with the experimental data \cite{PDG}.

 In the QCD sum rules for the heavy-light mesons, the Wilson coefficients of the vacuum condensates  at the operator product expansion side from different references differ from each other in one way or the other according to the   different approximations  \cite{QCDSRfD2,QCDSRfD5,WangJHEP,QCDSRfD1,QCDSR-mass}. In this article, we  recalculate the contributions of the   vacuum condensates up to dimension-6, including the one-loop corrections  to the quark condensates, and take into account the terms neglected in previous works,
then study the masses and decay constants of the pseudoscalar, scalar, vector and axial-vector  heavy-light mesons in a systematic way.

There have been many theoretical works on the decay constants of the heavy-light mesons, such as the QCD sum rules \cite{QCDSRfD2,QCDSRfD5,QCDSRfD4,QCDSRfD6,fD-Narison-2008,WangJHEP,QCDSRfD1,QCDSR-mass,WangCPL,SRBaker,QCDSRfDv1,QCDSRfD01,Aliev-pert,Eletsky-pert,QCDSR-3loop,SRfDEFT1,Narison1404}, the lattice QCD  \cite{LattfD1,LattfD2,LattfD3,LattfD-Dv1,LattfDv1,LattfDv2,LattfDv3,LattfD01,fB-exp-PDG}, the Bethe-Salpeter equation  \cite{BSEfD1,BSEfD2}, the relativistic potential model  \cite{RPMfD1,RPMfD2},  the field-correlator method  \cite{FCMfD1}, the light-front quark model  \cite{LFQMfD1,LFQMfDv1}, the chiral extrapolation \cite{Guo-fD}, the extended  chiral-quark model \cite{ChQM-fD}, the constituent quark model \cite{Khlopov}, etc.

The article is arranged as follows:  we derive the QCD sum rules for
the masses and decay constants of the heavy-light   mesons  in Sect.2;
in Sect.3, we present the numerical results and discussions; and Sect.4 is reserved for our
conclusions.

\section{QCD sum rules for  the heavy-light  mesons }
In the following, we write down  the two-point correlation functions
$\Pi_{0/5}(p)$ and $\Pi^{\mu\nu}_{V/A}(p)$ in the QCD sum rules,
\begin{eqnarray}
\Pi_{0/5}(p)&=&i\int d^4x e^{ip \cdot x} \langle 0|T\left\{J_{0/5}(x)J_{0/5}^{\dagger}(0)\right\}|0\rangle \, , \\
\Pi^{\mu\nu}_{V/A}(p)&=&i\int d^4x e^{ip \cdot x} \langle 0|T\left\{J^{\mu}_{V/A}(x)J^{\nu\dagger}_{V/A} (0)\right\}|0\rangle \, ,  \\
J_{0}(x)&=&\bar{Q}(x) q(x)  \, , \nonumber \\
J_{5}(x)&=&\bar{Q}(x)i\gamma_5 q(x)  \, , \nonumber \\
J_{V}^\mu(x)&=&\bar{Q}(x)\gamma^\mu q(x)  \, , \nonumber \\
J_{A}^\mu(x)&=&\bar{Q}(x)\gamma^\mu\gamma_5 q(x)  \, ,
\end{eqnarray}
where the currents $J_{5}(x)$, $J_{0}(x)$, $J_{V}^\mu(x)$ and $J_{A}^\mu(x)$ interpolate the  pseudoscalar, scalar, vector and axial-vector heavy-light mesons, respectively, $Q=c,b$ and $q=u,d,s$.
We can insert  a complete set of intermediate hadronic states with
the same quantum numbers as the current operators $J_{5}(x)$, $J_{0}(x)$, $J_{V}^\mu(x)$ and $J_{A}^\mu(x)$ into the
correlation functions $\Pi_{0/5}(p)$ and $\Pi^{\mu\nu}_{V/A}(p)$ to obtain the hadronic representation
\cite{SVZ79,Reinders85}. After isolating the ground state
contributions from the  pseudoscalar, scalar, vector and axial-vector heavy-light mesons, we get the following results,
\begin{eqnarray}
\Pi_0(p)&=&\frac{f_{S}^2m_{S}^2}{m_{S}^2-p^2} +\cdots\,  ,  \\
\Pi_5(p)&=&\frac{f_{P}^2m_{P}^4}{(m_Q+m_q)^2(m_{P}^2-p^2)} +\cdots\,  ,  \\
\Pi_{V/A}^{\mu\nu}(p)&=&\frac{f_{V/A}^2m_{V/A}^2}{m_{V/A}^2-p^2}\left( -g_{\mu\nu}+\frac{p^\mu p^\nu}{p^2}\right) +\cdots \nonumber\\
&=&\Pi_{V/A}(p)\left( -g_{\mu\nu}+\frac{p^\mu p^\nu}{p^2}\right) +\cdots \,  ,
\end{eqnarray}
where the  decay constants $f_{S/P/V/A}$ are defined by
\begin{eqnarray}
\langle 0|J_{0}(0)|S(p)\rangle&=&f_{S}m_{S}  \, , \nonumber\\
\langle 0|J_{5}(0)|P(p)\rangle&=&\frac{f_{P}m^2_{P}}{m_Q+m_q}  \, , \nonumber\\
\langle 0|J_{V/A}^\mu(0)|V/A(p)\rangle&=&f_{V/A}m_{V/A}\epsilon^\mu  \, ,
\end{eqnarray}
the $\epsilon^\mu$ are the polarization vectors of the vector and axial-vector mesons.

Now we carry out the operator product expansion at large space-like region $P^2=-p^2$.
The analytical expressions of the perturbative $\mathcal{O}(\alpha_s)$ corrections to the perturbative terms for  all the correlation functions  \cite{Aliev-pert,Eletsky-pert} and the semi-analytical
expressions of the perturbative $\mathcal{O}(\alpha_s^2)$ corrections to the perturbative terms for the pseudoscalar current's correlation functions  \cite{QCDSR-3loop}  are available now. We take  into account those  analytical and semi-analytical expressions directly \cite{Aliev-pert,Eletsky-pert,QCDSR-3loop}; and recalculate the  contributions  of the vacuum condensates, i.e.  we calculate the Feynman diagrams shown in Figs.1-5, where the solid and dashed lines denote the light and heavy quark lines, respectively, the wave line denotes the gluon line. In  calculating the diagrams in Fig.2, we correct  the minor errors in Ref.\cite{WangJHEP}, where the  quark condensate $\frac{\langle \bar{q}q\rangle}{12}$ in the full light-quark propagators is replaced with  $\frac{\langle \bar{q}q\rangle}{3D}$, the $D$ is the dimension of the space-time. A minor error occurs when there exist divergences, such a step should be deleted, i.e. the quark condensate $\frac{\langle \bar{q}q\rangle}{12}$ survives in the $D$-dimension.  In Ref.\cite{WangNM}, we correct the minor errors and improve the calculations,  and obtain the correct expressions. Furthermore, we obtain the perturbative $\mathcal{O}(\alpha_s)$ corrections to the quark condensate terms  for the vector and axial-vector currents.

\begin{figure}
 \centering
 \includegraphics[totalheight=2.5cm,width=4cm]{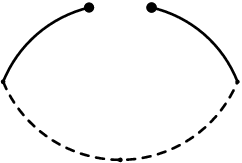}
    \caption{The diagram contributes to   the quark condensate $\langle\bar{q}q\rangle$. }
\end{figure}
\begin{figure}
 \centering
 \includegraphics[totalheight=6cm,width=14cm]{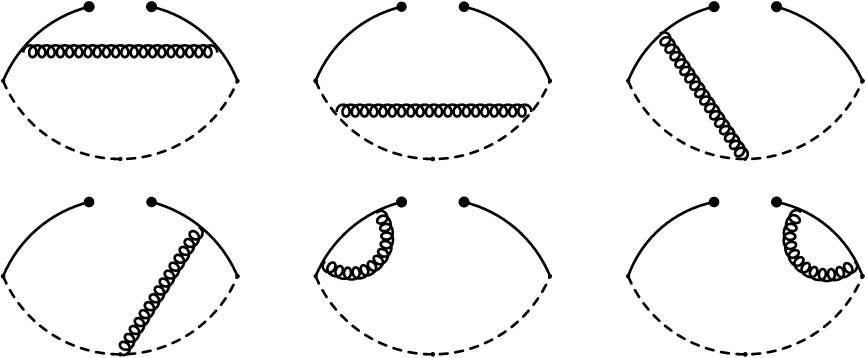}
    \caption{The perturbative $\mathcal{O}(\alpha_s)$ corrections to the quark condensate $\langle\bar{q}q\rangle$. }
\end{figure}
\begin{figure}
 \centering
 \includegraphics[totalheight=3cm,width=14cm]{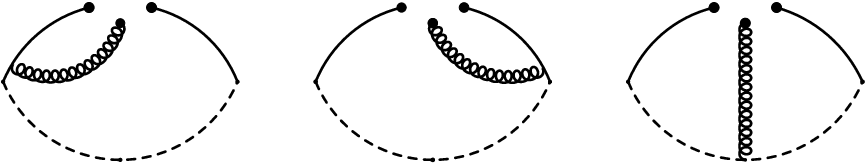}
    \caption{The diagrams contribute to the mixed condensate $\langle\bar{q}g_s \sigma G q\rangle$. }
\end{figure}
\begin{figure}
 \centering
 \includegraphics[totalheight=6cm,width=14cm]{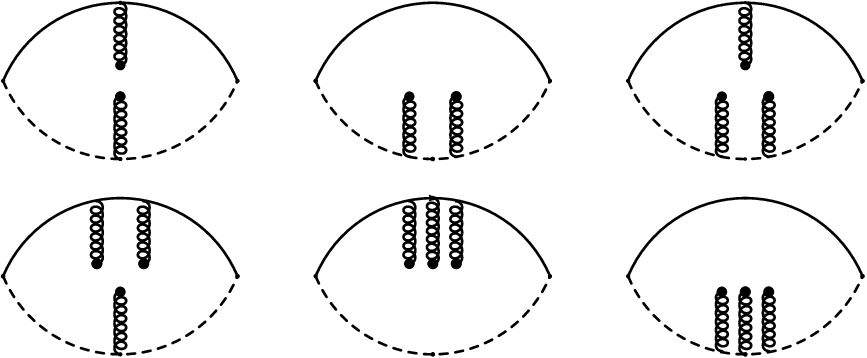}
    \caption{The diagrams contribute to the gluon condensate $\langle \frac{\alpha_sGG}{\pi}\rangle$ and three-gluon condensate $\langle g_s^3 GGG\rangle$. }
\end{figure}
\begin{figure}
 \centering
 \includegraphics[totalheight=3cm,width=14cm]{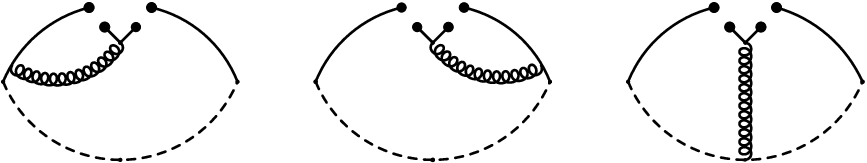}
    \caption{The diagrams contribute to the four-quark condensate $\langle\bar{q} q\rangle^2$. }
\end{figure}

 Once analytical expressions of the QCD spectral densities are obtained,   then we can take the
quark-hadron duality below the continuum thresholds and perform the Borel transforms  with respect to the variable
$P^2=-p^2$ to obtain the QCD sum rules,

\begin{eqnarray}
\frac{f_P^2m_P^4}{(m_Q+m_q)^2}\exp\left(-\frac{m_P^2}{T^2} \right) &=&B_T \Pi_5\, , \\
f_S^2m_S^2\exp\left(-\frac{m_S^2}{T^2} \right) &=&B_T \Pi_0\, , \\
f_V^2m_V^2\exp\left(-\frac{m_V^2}{T^2} \right) &=&B_T \Pi_V\, , \\
f_A^2m_A^2\exp\left(-\frac{m_A^2}{T^2} \right) &=&B_T \Pi_A\, ,
\end{eqnarray}
where
\begin{eqnarray}
B_T \Pi_5&=&B_T\Pi_5^0+B_T\Pi_5^3+B_T \Pi_5^4+B_T \Pi_5^5+B_T \Pi_5^6\, ,\nonumber\\
B_T \Pi_0&=&B_T\Pi_5|_{m_Q \to -m_Q}\, ,\\
B_T \Pi_V&=&B_T\Pi_V^0+B_T\Pi_V^3+B_T \Pi_V^4+B_T \Pi_V^5+B_T \Pi_V^6\, ,\nonumber\\
B_T \Pi_A&=&B_T\Pi_V|_{m_Q \to -m_Q}\, ,
\end{eqnarray}

\begin{eqnarray}
B_T\Pi_5^0&=& \frac{3}{8\pi^2} \int_{m_Q^2}^{s_0} ds s\left(1-\frac{m_Q^2}{s}\right)^2\left\{1+\frac{2m_q m_Q}{s-m_Q^2}+\frac{4\alpha_s}{3\pi} R_5\left(\frac{m_Q^2}{s}\right)\right\} \exp\left(-\frac{s}{T^2}\right) \, ,
\\
B_T\Pi_5^3&=& -m_Q\langle\bar{q}q\rangle\left\{1+\frac{\alpha_s}{\pi} \left[ 6-\frac{4m_Q^2}{3T^2}-\frac{2}{3}\left( 1-\frac{m_Q^2}{T^2}\right)\log\frac{m_Q^2}{\mu^2}-2\Gamma\left(0,\frac{m_Q^2}{T^2}\right)\exp\left( \frac{m_Q^2}{T^2}\right) \right]\right\}\nonumber\\
&&\exp\left(- \frac{m_Q^2}{T^2}\right) +\frac{m_q\langle\bar{q}q\rangle}{2}\left(1+\frac{m_Q^2}{T^2}\right)\exp\left(-\frac{m_Q^2}{T^2}\right)\, ,
\\
B_T\Pi_5^4&=&\frac{1}{12}\langle \frac{\alpha_sGG}{\pi}\rangle\exp\left(-\frac{m_Q^2}{T^2}\right)\, ,
 \\
B_T\Pi_5^5&=& -\left\{\frac{m_Q\langle\bar{q}g_s\sigma Gq\rangle}{2T^2}\left(1-\frac{m_Q^2}{2T^2}\right)+\frac{m_q m_Q^4\langle\bar{q}g_s\sigma Gq\rangle}{12T^6}\right\}\exp\left(-\frac{m_Q^2}{T^2}\right) \, ,
\\
  B_T\Pi_5^6&=&-\frac{16\pi\alpha_s\langle\bar{q}q\rangle^2 }{27T^2}\left(1+\frac{m_Q^2}{2T^2}
 -\frac{m_Q^4}{12T^4}\right)\exp\left(-\frac{m_Q^2}{T^2}\right)\nonumber\\
 && +\frac{\langle g_s^3 GGG\rangle}{\pi^2}\left\{ \frac{5}{192T^2}+ \frac{1}{768m_Q^2} +\frac{5m_Q^2}{1536T^4}-\frac{m_Q^4}{768T^6}-\left(\frac{m_Q^2}{128T^4}+\frac{m_Q^4}{384T^6} \right)\log \frac{m_Q^2\mu^2}{T^4}\right\}\nonumber\\
 &&\exp\left(-\frac{m_Q^2}{T^2}\right)  \, ,
\end{eqnarray}
\begin{eqnarray}
B_T\Pi_V^0&=& \frac{1}{8\pi^2} \int_{m_Q^2}^{s_0} ds s\left(1-\frac{m_Q^2}{s}\right)^2\left(2+\frac{m_Q^2}{s} \right)\left\{1+\frac{6sm_q m_Q}{(s-m_Q^2)(2s+m_Q^2)}+\frac{4\alpha_s}{3\pi} R_V\left(\frac{m_Q^2}{s}\right)\right\} \nonumber\\
&&\exp\left(-\frac{s}{T^2}\right) \, ,
\\
B_T\Pi_V^3&=&-m_Q\langle\bar{q}q\rangle\left\{1+\frac{\alpha_s}{\pi} \left[ \frac{8}{3}-\frac{4m_Q^2}{3T^2} +\frac{2}{3}\left( 2+\frac{m_Q^2}{T^2}\right)\log\frac{m_Q^2}{\mu^2}-\frac{2m_Q^2}{3T^2}\Gamma\left(0,\frac{m_Q^2}{T^2}\right)\exp\left( \frac{m_Q^2}{T^2}\right) \right]\right\}\nonumber\\
&&\exp\left(- \frac{m_Q^2}{T^2}\right)+\frac{m_q m_Q^2\langle\bar{q}q\rangle}{2T^2} \exp\left(-\frac{m_Q^2}{T^2}\right) \, ,
\\
B_T\Pi_V^4&=&-\frac{1}{12}\langle \frac{\alpha_sGG}{\pi}\rangle\exp\left(-\frac{m_Q^2}{T^2}\right)\, ,
 \\
B_T\Pi_V^5&=& \left\{\frac{m_Q^3\langle\bar{q}g_s\sigma Gq\rangle}{4T^4}+\frac{m_q \langle\bar{q}g_s\sigma Gq\rangle}{12T^2}\left( 1+\frac{m_Q^2}{T^2}-\frac{m_Q^4}{T^4} \right)\right\}\exp\left(-\frac{m_Q^2}{T^2}\right) \, ,
\\
 B_T\Pi_V^6&=&-\frac{20\pi\alpha_s\langle\bar{q}q\rangle^2 }{81T^2}\left(1+\frac{m_Q^2}{T^2}
 -\frac{m_Q^4}{5T^4}\right)\exp\left(-\frac{m_Q^2}{T^2}\right)+\frac{\langle g_s^3 GGG\rangle}{\pi^2} \left\{ -\frac{1}{1152T^2}+ \frac{1}{1152m_Q^2} \right.\nonumber\\
 &&\left.+\frac{m_Q^2}{768T^4}\left(1-\frac{m_Q^2}{T^2} \right)+\frac{55m_Q^2}{4608T^4}+\frac{1}{192T^2}\left(1+\frac{m_Q^2}{T^2}-\frac{m_Q^4}{2T^4} \right)\log \frac{m_Q^2\mu^2}{T^4}\right\}\exp\left(-\frac{m_Q^2}{T^2}\right)  \, , \nonumber\\
\end{eqnarray}
\begin{eqnarray}
R_5(x)&=&\frac{9}{4}+2{\rm Li}_2(x)+{ \log}x\,{\log}(1-x)-\frac{3}{2}\,{\log}\frac{1-x}{x}-{\log}(1-x)+x\,{\log}\frac{1-x}{x}
-\frac{x}{1-x}{\log}x \, , \nonumber\\
R_V(x)&=&\frac{13}{4}+2{\rm Li}_2(x)+{ \log}x\,{\log}(1-x)-\frac{3}{2}\,{\log}\frac{1-x}{x}-{\log}(1-x)+x\,{\log}\frac{1-x}{x}
-\frac{x}{1-x}{\log}x   \nonumber\\
&&+\frac{(3+x)(1-x)}{2+x}\log \frac{1-x}{x}-\frac{2x}{(2+x)(1-x)^2}\log x -\frac{5+2x}{2+x}-\frac{2x}{(2+x)(1-x)} \, ,
\end{eqnarray}
\begin{eqnarray}
\Gamma(0,x)&=&e^{-x}\int_0^\infty dt \frac{1}{t+x}e^{-t} \, , \nonumber\\
{\rm Li}_2(x)&=&-\int_0^x dt \frac{1}{t} \log(1-t)\, ,
\end{eqnarray}
and the $s_0$ are the continuum threshold parameters.
The perturbative $\mathcal{O}(\alpha_s)$ corrections $R_5(x)$ and $R_V(x)$ are taken from Refs.\cite{Aliev-pert,Eletsky-pert}.
We can also take into account the
semi-analytical perturbative  $\mathcal{O}(\alpha_s^2)$ corrections to the perturbative terms for the $B_T\Pi_5^0$ ,
\begin{eqnarray}
\frac{1}{8\pi^2}\left(\frac{\alpha_s}{\pi}\right)^2\int_{m_c^2}^{s_0}ds \left\{ \frac{16}{9}\,{\rm R2sFF}[v]+4\,{\rm R2sFA}[v]+\frac{2n_l}{3}\, {\rm R2sFL}[v]
+\frac{2}{3}\,{\rm R2sFH} [v]\right\} \exp\left(-\frac{s}{T^2}\right)\, ,
\end{eqnarray}
where the ${\rm R2sFF} [v]$, ${\rm R2sFA} [v]$, ${\rm R2sFL} [v]$ and
${\rm R2sFH} [v]$ with the variable  $v=\left(1-\frac{m_c^2}{s} \right)/\left(1+\frac{m_c^2}{s} \right)$ are mathematical functions defined at the energy-scale of the pole mass $\mu=m_c$, here the $n_l$ counts the number of massless quarks \cite{QCDSR-3loop}.

 We can derive Eqs.(9-12) with respect to $1/T^2$, then eliminate the decay constants $f_{S/P/V/A}$ to obtain the QCD sum rules for the masses.
\begin{eqnarray}
m^2_{S/P/V/A}&=&\frac{-\frac{d}{d(1/T^2)}B_T\Pi_{0/5/V/A}}{B_T\Pi_{0/5/V/A}}\, .
\end{eqnarray}
Once the masses $m_{S/P/V/A}$ are obtained, we can take them as input parameters and obtain the decay constants from the QCD sum rules in Eqs.(9-12).

In the case of the light-quark currents, the perturbative $\mathcal{O}(\alpha_s)$ corrections to the perturbative terms amount to multiplying the  factors
$1+\frac{11}{3}\frac{\alpha_s}{\pi}\approx 1+3.67\frac{\alpha_s}{\pi}$ and $ 1+\frac{\alpha_s}{\pi}$ to the perturbative terms in the correlation functions for  the  pseudoscalar (scalar) and vector (axial-vector) currents, respectively \cite{Reinders85}. In the present case, if we take the approximation $\mu^2=m_c^2=T^2$, the perturbative  $\mathcal{O}(\alpha_s)$ corrections to the quark condensate terms amount to multiplying the factors $1+3.47\frac{\alpha_s}{\pi}$ and $1+0.94\frac{\alpha_s}{\pi}$ to the quark condensate terms in the correlation functions for the  pseudoscalar (scalar) and vector (axial-vector) currents, respectively. The analogous $\mathcal{O}(\alpha_s)$ corrections  indicate that the present calculations are reliable.

\section{Numerical results and discussions}
In the heavy quark limit, the heavy-light  mesons
$Q{\bar q}$  can be  classified in doublets according to the total
angular momentum of the light antiquark ${\vec s}_\ell$,
${\vec s}_\ell= {\vec s}_{\bar q}+{\vec L} $, where the ${\vec
s}_{\bar q}$ and ${\vec L}$ are the spin and orbital angular momentum of the light antiquark,  respectively.   The spin doublets  $(D,D^*)$, $(D_s,D_s^*)$, $(D_0^*(2400),D_1(2430))$, $(D_{s0}^*(2317),D_{s1}(2460))$, $(B,B^*)$, $(B_s,B_s^*)$ have been observed,   the  masses   are
$m_{D^\pm}=(1869.5 \pm 0.4)\,\rm{MeV}$,
$m_{D^0}=(1864.84 \pm 0.07)\,\rm{MeV}$,
$m_{D^{*}(2010)^\pm}=(2010.26\pm0.07)\,\rm{MeV}$,
$m_{D^{*}(2007)^0}=(2006.96\pm0.10)\,\rm{MeV}$,
$m_{D_0^{*}(2400)^0}=(2318\pm29)\,\rm{MeV}$,
$m_{D_0^{*}(2400)^\pm}=(2403\pm14\pm35)\,\rm{MeV}$,
$m_{D_1(2430)^0}=(2427\pm26\pm25)\,\rm{MeV}$,
$m_{D_s^\pm}=(1969.0 \pm 1.4)\,\rm{MeV}$,
$m_{D_s^{*}(2112)^\pm}=(2112.1\pm0.4)\,\rm{MeV}$,
$m_{D_{s0}^{*}(2317)^\pm}=(2318.0\pm1.0)\,\rm{MeV}$,
$m_{D_{s1}(2460)^\pm}=(2459.6\pm0.9)\,\rm{MeV}$,
$m_{B^\pm}=(5279.25\pm0.26)\,\rm{MeV}$,
 $m_{B^0}=(5279.55\pm0.26)\,\rm{MeV}$,
 $m_{B^*}=(5325.2\pm0.4)\,\rm{MeV}$,
  $m_{B_s}=(5366.7 \pm 0.4)\,\rm{MeV}$,
  $m_{B_s^*}=(5415.8\pm1.5)\,\rm{MeV}$  from the    Particle Data Group
\cite{PDG}. The spin doublets $(B^*_{0}, B_{1})$ and  $(B^*_{s0}, B_{s1})$ have not been observed yet.  The doublet    $(D(2550), D(2600))$ or $(D_J(2580), D_J^*(2650))$ is tentatively identified as the first radial excited state of the doublet
$(D,D^*)$, the doublet $(?,D_{s1}^*(2700))$ is tentatively identified as the first radial excited state of the doublet
$(D_s,D^*_s(2112))$ \cite{D2550}.

We take the values  $\sqrt{s_0}=m_{\rm gr}+(0.4-0.8)\,\rm{GeV}$ as guides,  here the gr denotes the ground states, and search for the optimal threshold parameters $s_0$ and Borel parameters $T^2$ to satisfy the  following criteria:

$\bullet$ Pole dominance at the phenomenological side;

$\bullet$ Convergence of the operator product expansion;

$\bullet$ Appearance of the Borel platforms;

$\bullet$ Reappearance  of experimental values of the ground state heavy meson masses.

The contributions of the ground states can be fully taken into account
by choosing the   threshold parameters $\sqrt{s_0}=m_{\rm gr}+(0.4-0.8)\,\rm{GeV}$.
The contaminations of the excited states are very small if there are some contaminations,  we expect that the couplings of
 the  currents  to the excited states are more weak than that to the ground states. For example, the decay constants of the pseudoscalar mesons $\pi(140)$ and $\pi(1800)$ have the hierarchy $f_{\pi(1300)}\ll f_{\pi(140)}$ from the lattice QCD \cite{Latt-pion},  the QCD sum rules \cite{QCDSR-pion},  or from the experimental data \cite{pion-exp}.

The vacuum condensates are taken to be the standard values
$\langle\bar{u}u \rangle=\langle\bar{d}d \rangle=-(0.24\pm 0.01\, \rm{GeV})^3$,  $\langle\bar{s}s \rangle=(0.8\pm0.1)\langle\bar{u}u \rangle$,
$\langle\bar{q}g_s\sigma G q \rangle=m_0^2\langle \bar{q}q \rangle$,
$m_0^2=(0.8 \pm 0.1)\,\rm{GeV}^2$, $\langle \frac{\alpha_s
GG}{\pi}\rangle=(0.33\,\rm{GeV})^4 $, $\langle g_s^3 GGG\rangle=0.045\,\rm{GeV}^6$    at the energy scale  $\mu=1\, \rm{GeV}$
\cite{SVZ79,Reinders85}.
The quark condensates and mixed quark condensates  evolve with the   renormalization group equation,
$\langle\bar{q}q \rangle(\mu)=\langle\bar{q}q \rangle(Q)\left[\frac{\alpha_{s}(Q)}{\alpha_{s}(\mu)}\right]^{\frac{4}{9}}$ and
 $\langle\bar{q}g_s \sigma Gq \rangle(\mu)=\langle\bar{q}g_s \sigma Gq \rangle(Q)\left[\frac{\alpha_{s}(Q)}{\alpha_{s}(\mu)}\right]^{\frac{2}{27}}$.

In the article, we take the $\overline{MS}$ masses $m_{b}(m_b)=(4.18\pm0.03)\,\rm{GeV}$, $m_{c}(m_c)=(1.275\pm0.025)\,\rm{GeV}$ and $m_s(\mu=2\,\rm{GeV})=(0.095\pm0.005)\,\rm{GeV}$
 from the Particle Data Group \cite{PDG}, and take into account
the energy-scale dependence of  the $\overline{MS}$ masses from the renormalization group equation,
\begin{eqnarray}
m_b(\mu)&=&m_b(m_b)\left[\frac{\alpha_{s}(\mu)}{\alpha_{s}(m_b)}\right]^{\frac{12}{23}} \, ,\nonumber\\
m_c(\mu)&=&m_c(m_c)\left[\frac{\alpha_{s}(\mu)}{\alpha_{s}(m_c)}\right]^{\frac{12}{25}} \, ,\nonumber\\
m_s(\mu)&=&m_s({\rm 2GeV} )\left[\frac{\alpha_{s}(\mu)}{\alpha_{s}({\rm 2GeV})}\right]^{\frac{4}{9}} \, ,\nonumber\\
m_{u/d}(\mu)&=&m_{u/d}({\rm 1GeV} )\left[\frac{\alpha_{s}(\mu)}{\alpha_{s}({\rm 1GeV})}\right]^{\frac{4}{9}} \, ,\nonumber\\
\alpha_s(\mu)&=&\frac{1}{b_0t}\left[1-\frac{b_1}{b_0^2}\frac{\log t}{t} +\frac{b_1^2(\log^2{t}-\log{t}-1)+b_0b_2}{b_0^4t^2}\right]\, ,
\end{eqnarray}
  where $t=\log \frac{\mu^2}{\Lambda^2}$, $b_0=\frac{33-2n_f}{12\pi}$, $b_1=\frac{153-19n_f}{24\pi^2}$, $b_2=\frac{2857-\frac{5033}{9}n_f+\frac{325}{27}n_f^2}{128\pi^3}$,  $\Lambda=213\,\rm{MeV}$, $296\,\rm{MeV}$  and  $339\,\rm{MeV}$ for the flavors  $n_f=5$, $4$ and $3$, respectively  \cite{PDG}.
 Furthermore, we obtain the values $m_u=m_d=6\,\rm{MeV}$ from the Gell-Mann-Oakes-Renner relation at the energy scale $\mu=1\,\rm{GeV}$.

In this article, we choose the $\overline{MS}$ masses by setting $m=m(\mu)$ and take the perturbative $\mathcal{O}(\alpha_s)$ corrections  to the perturbative terms. In other words, we take  the $R_{5/V}\left(\frac{m_Q^2}{s}\right)$ only.
In calculations, we take  $n_f=3$ and $\mu_{D/D^*}=\sqrt{m_D^2-m_c^2}\approx 1\,\rm{GeV}$ for the S-wave mesons $D$ and $D^*$;   $n_f=4$ and $\mu_{B/B^*}=\sqrt{m_B^2-m_b^2}\approx 2.5\,\rm{GeV}$ for the S-wave mesons $B$ and $B^*$. If we count  the contribution of the additional P-wave as  $0.5\,\rm{GeV}$, then $\mu_{D_0^*/D_1}=1.5\,\rm{GeV}$ and $\mu_{B_0^*/B_1}=3.0\,\rm{GeV}$. On the other hand, we take into account the $SU(3)$ breaking effect, which is supposed  to be  $100\,\rm{MeV}$ for the light quarks, then $\mu_{D_s/D_s^*}=1.1\,\rm{GeV}$,  $\mu_{B_s/B_s^*}=2.6\,\rm{GeV}$, $\mu_{D_{s0}^*/D_{s1}}=1.6\,\rm{GeV}$ and $\mu_{B_{s0}^*/B_{s1}}=3.1\,\rm{GeV}$. Those energy scales work well.

The  continuum threshold parameters, Borel parameters, pole contributions are shown explicitly in Table 1. From Table 1, we can see that the pole dominance can be satisfied. On the other hand, the dominant contributions come from the perturbative terms and the quark condensate terms, so we expect to obtain reliable predictions.

After taking into account the uncertainties of the input parameters, we obtain the values of the masses and decay constants of the heavy-light mesons, which are shown in Figs.6-9 and Table 1. From the figures, we can see that the masses and decays constants are rather stable with variations of the Boral parameters $T^2$, the predictions are reasonable.

\begin{figure}
 \centering
 \includegraphics[totalheight=5cm,width=6cm]{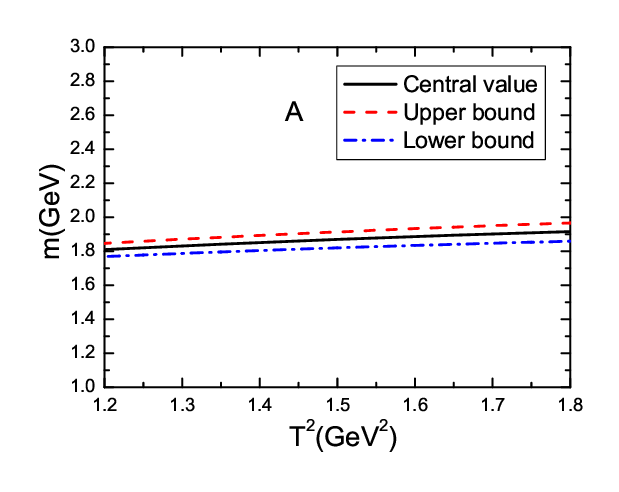}
 \includegraphics[totalheight=5cm,width=6cm]{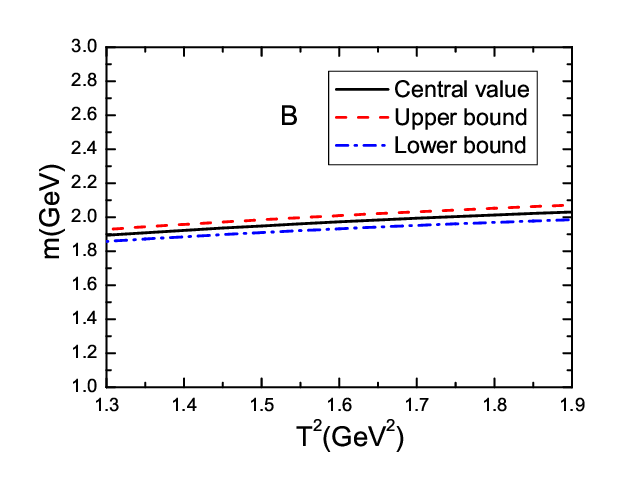}
 \includegraphics[totalheight=5cm,width=6cm]{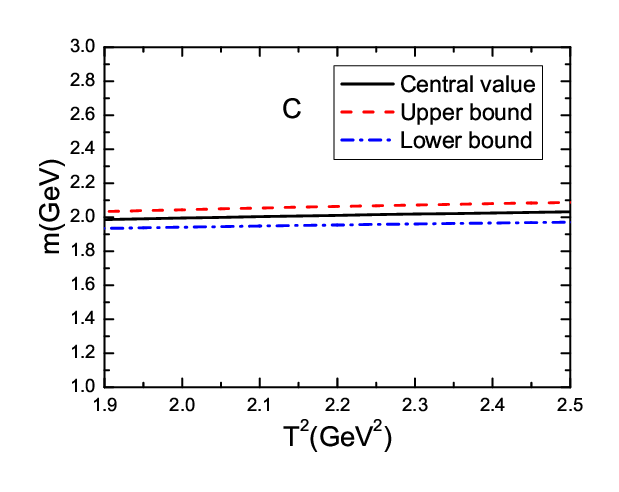}
 \includegraphics[totalheight=5cm,width=6cm]{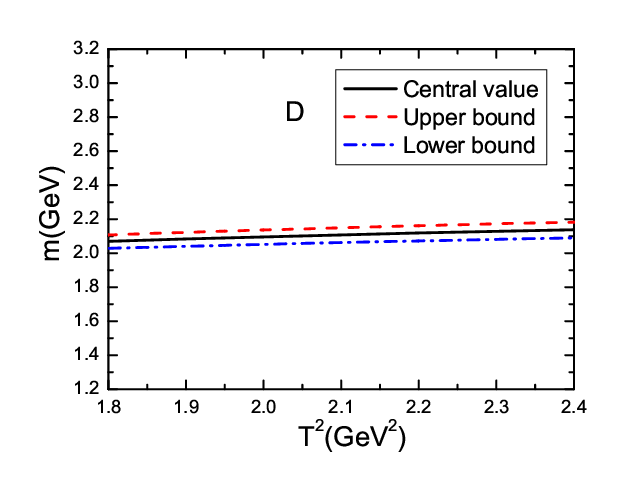}
 \includegraphics[totalheight=5cm,width=6cm]{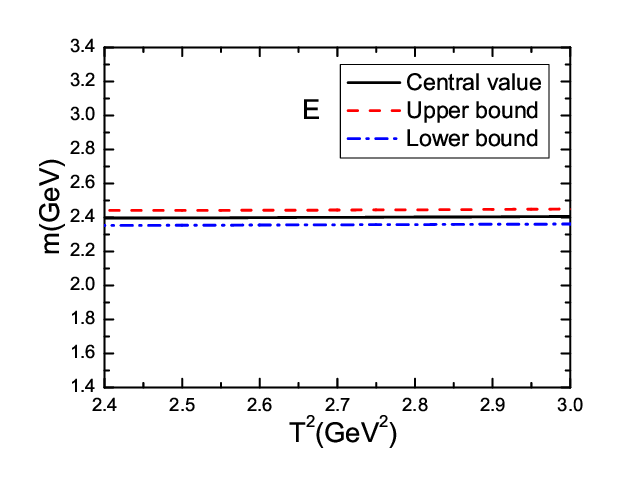}
 \includegraphics[totalheight=5cm,width=6cm]{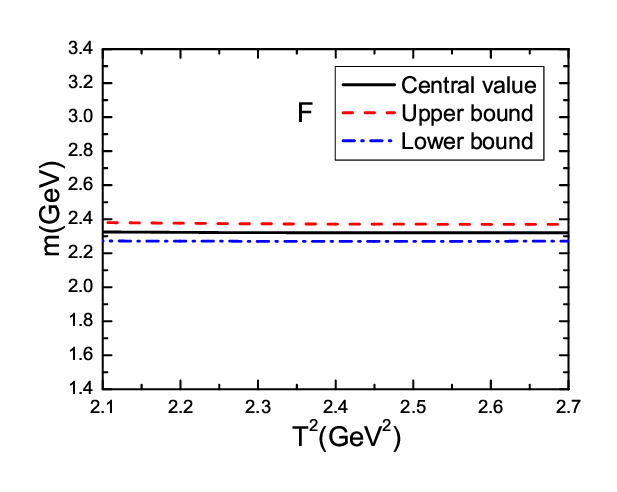}
 \includegraphics[totalheight=5cm,width=6cm]{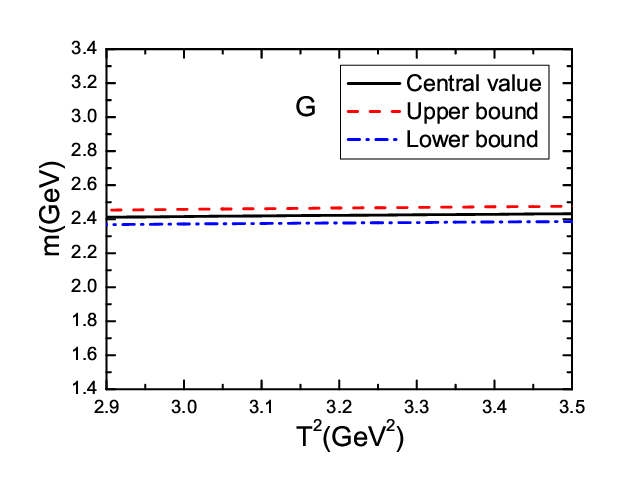}
 \includegraphics[totalheight=5cm,width=6cm]{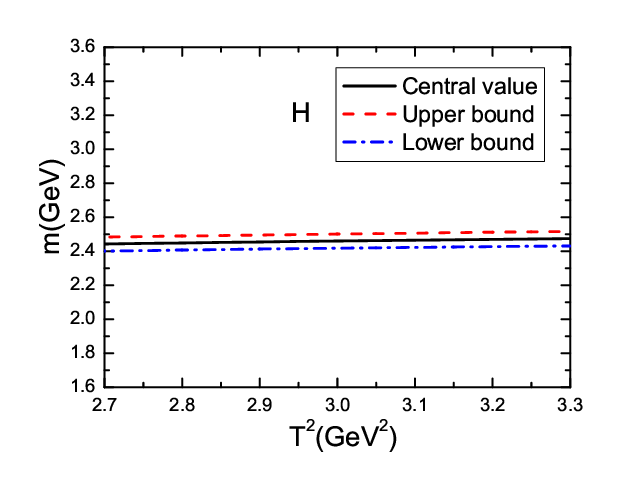}
        \caption{ The masses of the charmed mesons with variations of the Borel parameters $T^2$, the $A$, $B$, $C$, $D$, $E$, $F$, $G$ and $H$ denote  the mesons $D$, $D_s$, $D^*$, $D^*_s$, $D_0^*$, $D_{s0}^*$, $D_1$ and $D_{s1}$, respectively.}
\end{figure}

\begin{figure}
 \centering
 \includegraphics[totalheight=5cm,width=6cm]{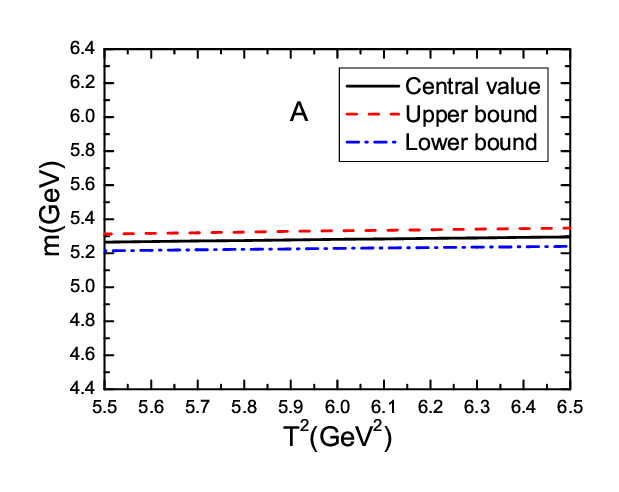}
 \includegraphics[totalheight=5cm,width=6cm]{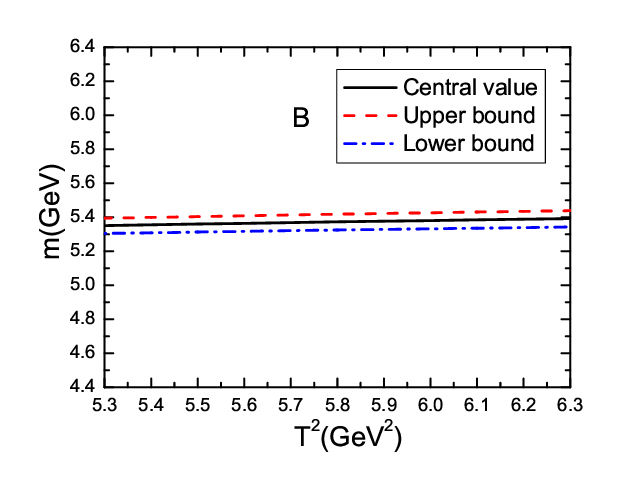}
 \includegraphics[totalheight=5cm,width=6cm]{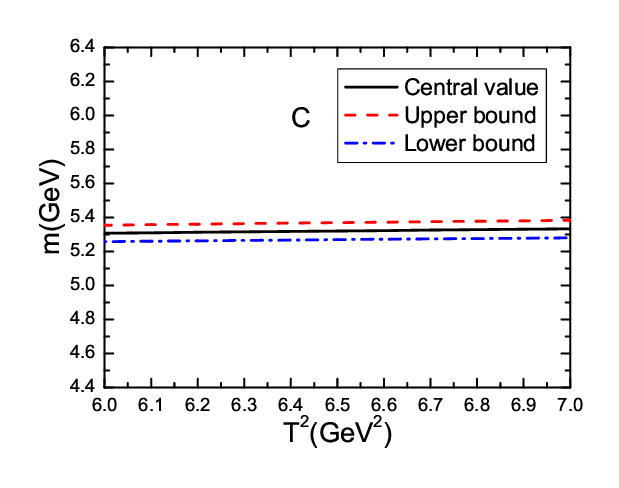}
 \includegraphics[totalheight=5cm,width=6cm]{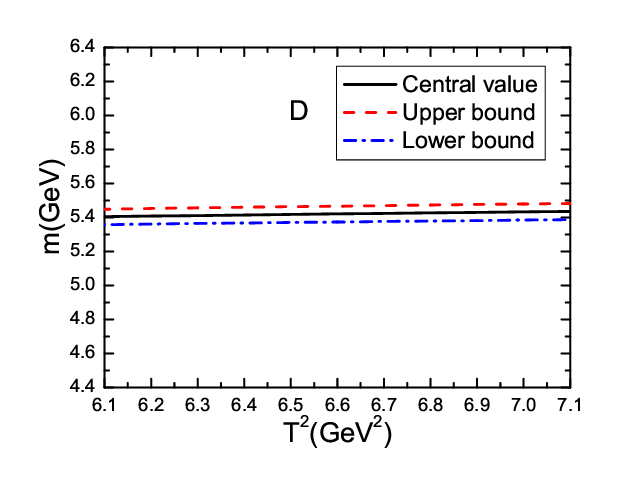}
 \includegraphics[totalheight=5cm,width=6cm]{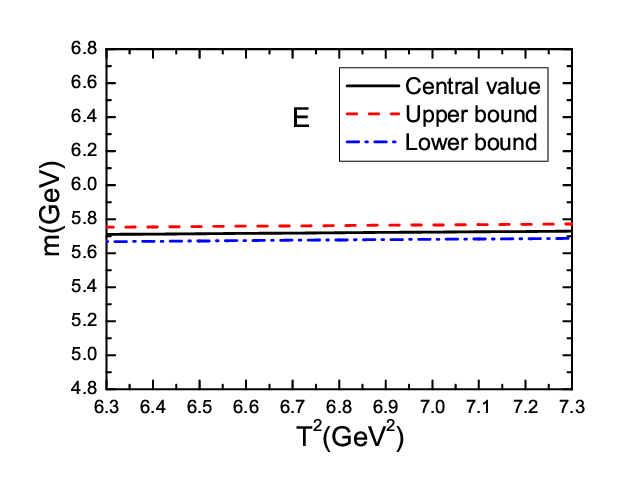}
 \includegraphics[totalheight=5cm,width=6cm]{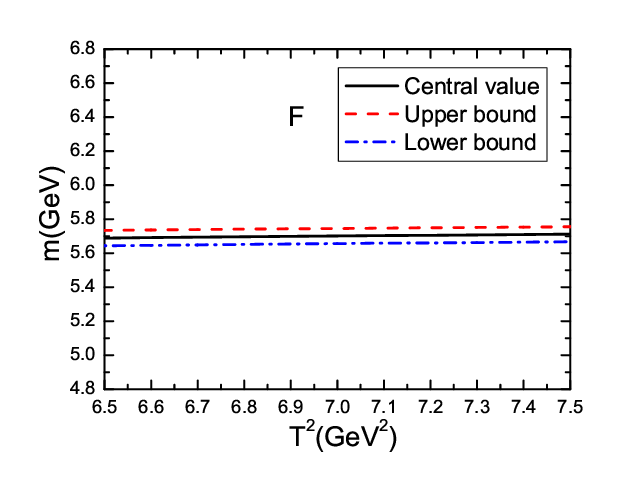}
 \includegraphics[totalheight=5cm,width=6cm]{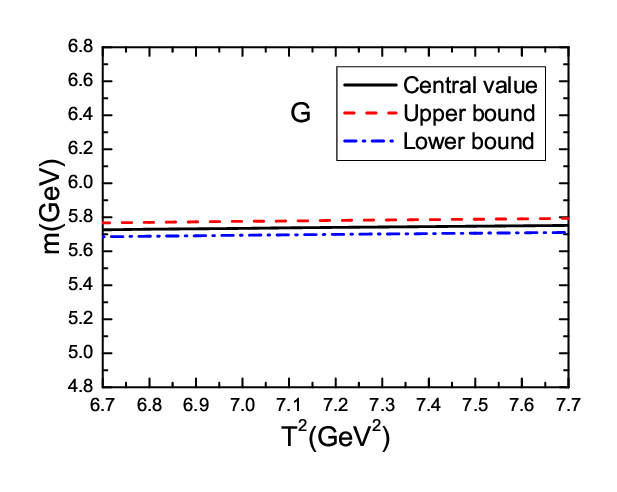}
 \includegraphics[totalheight=5cm,width=6cm]{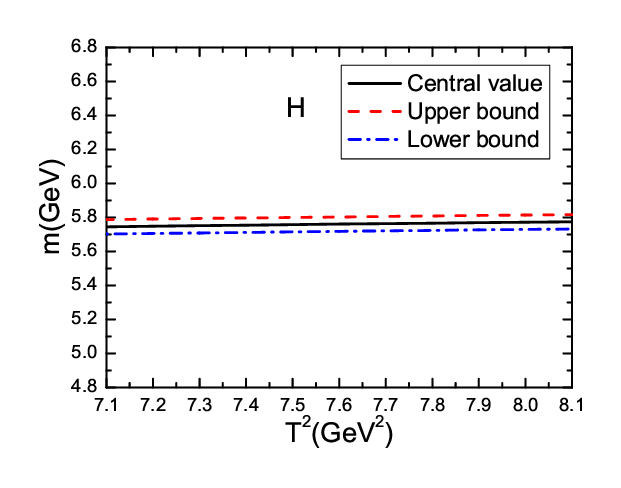}
        \caption{  The masses of the bottom mesons with variations of the Borel parameters $T^2$, the $A$, $B$, $C$, $D$, $E$, $F$, $G$ and $H$ denote  the mesons $B$, $B_s$, $B^*$, $B^*_s$, $B_0^*$, $B_{s0}^*$, $B_1$ and $B_{s1}$, respectively.}
\end{figure}

\begin{figure}
 \centering
 \includegraphics[totalheight=5cm,width=6cm]{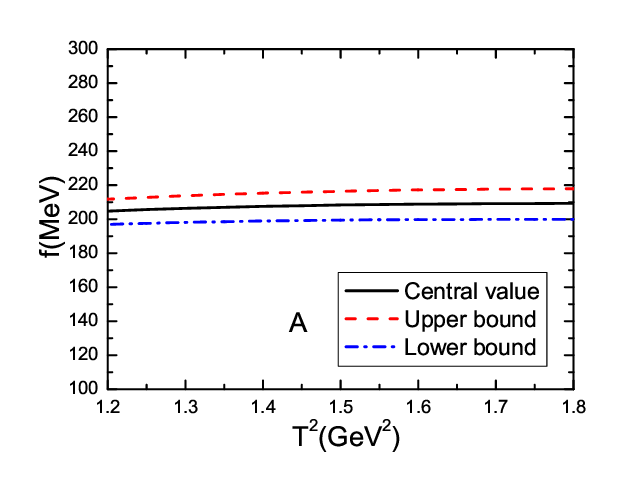}
 \includegraphics[totalheight=5cm,width=6cm]{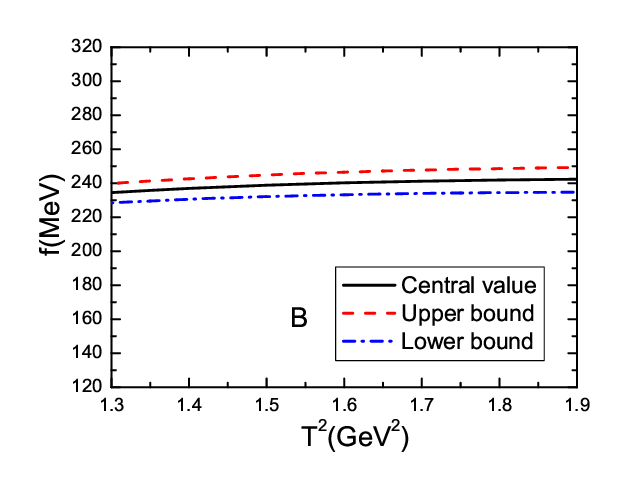}
 \includegraphics[totalheight=5cm,width=6cm]{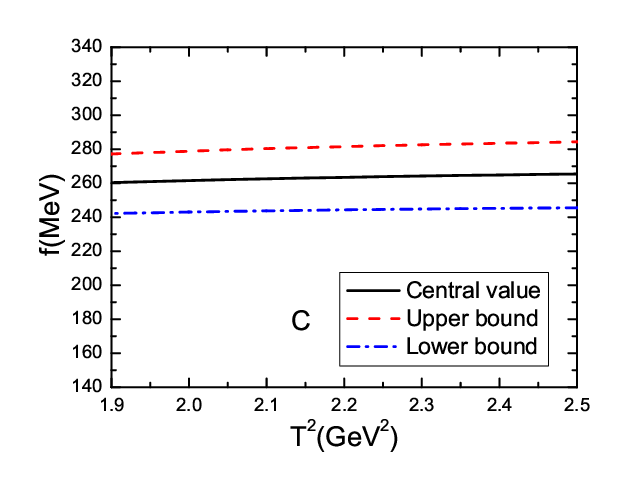}
 \includegraphics[totalheight=5cm,width=6cm]{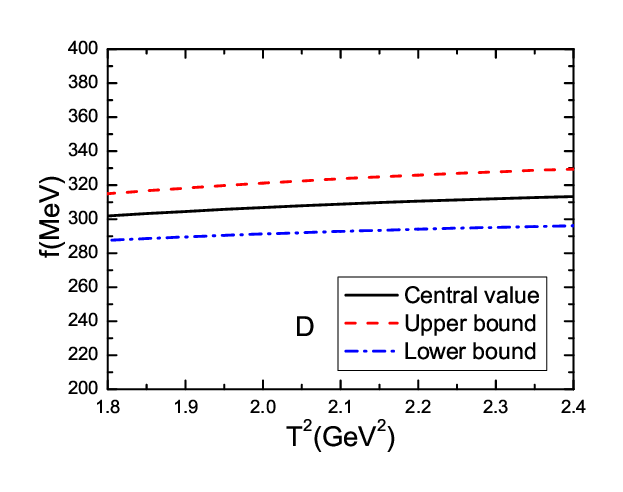}
 \includegraphics[totalheight=5cm,width=6cm]{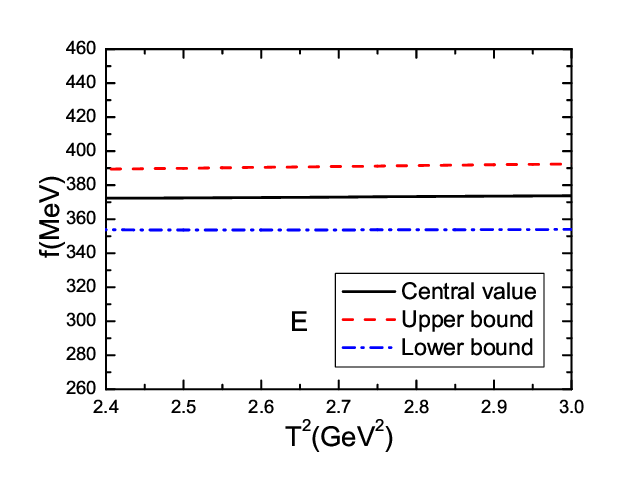}
 \includegraphics[totalheight=5cm,width=6cm]{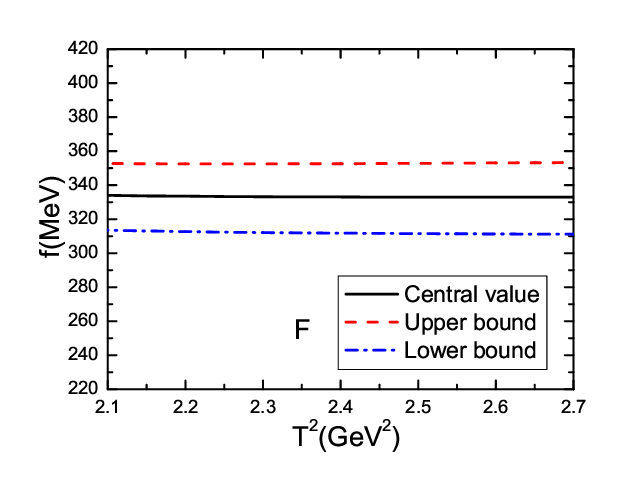}
 \includegraphics[totalheight=5cm,width=6cm]{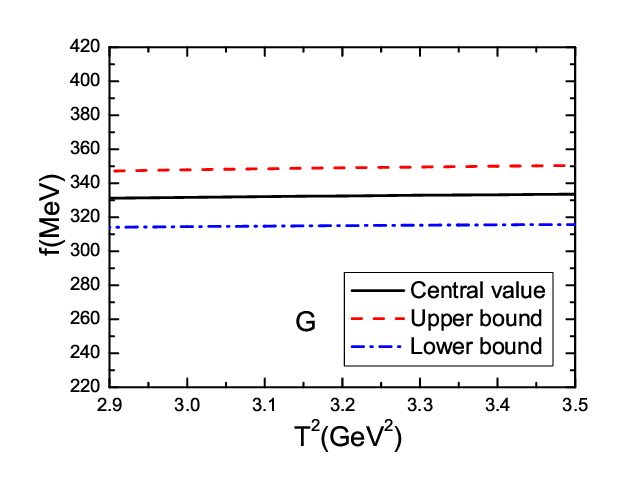}
 \includegraphics[totalheight=5cm,width=6cm]{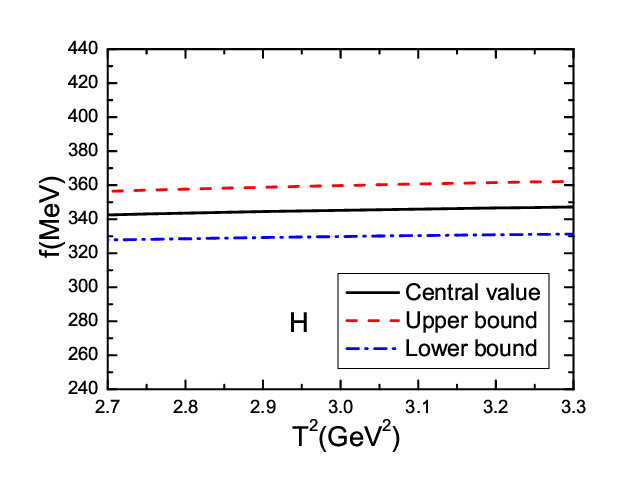}
        \caption{  The decay constants of the charmed mesons with variations of the Borel parameters $T^2$, the $A$, $B$, $C$, $D$, $E$, $F$, $G$ and $H$ denote  the mesons $D$, $D_s$, $D^*$, $D^*_s$, $D_0^*$, $D_{s0}^*$, $D_1$ and $D_{s1}$, respectively.}
\end{figure}

\begin{figure}
 \centering
 \includegraphics[totalheight=5cm,width=6cm]{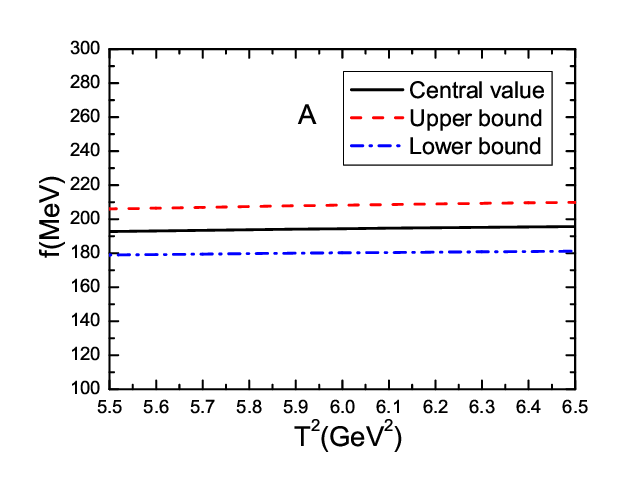}
 \includegraphics[totalheight=5cm,width=6cm]{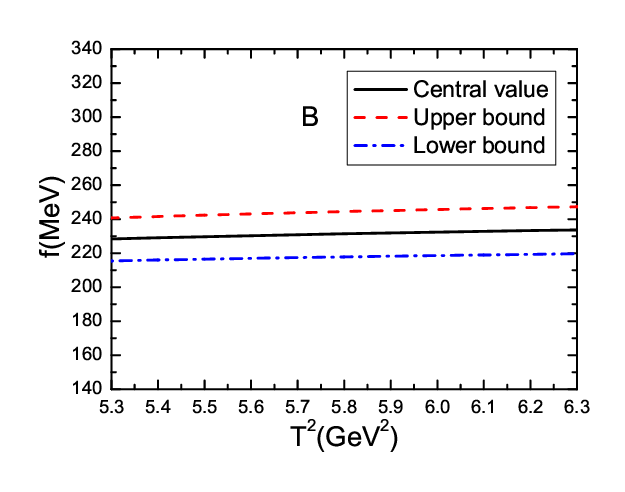}
 \includegraphics[totalheight=5cm,width=6cm]{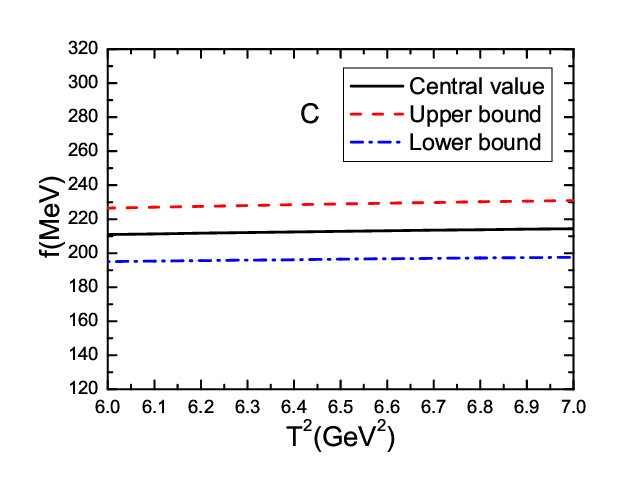}
 \includegraphics[totalheight=5cm,width=6cm]{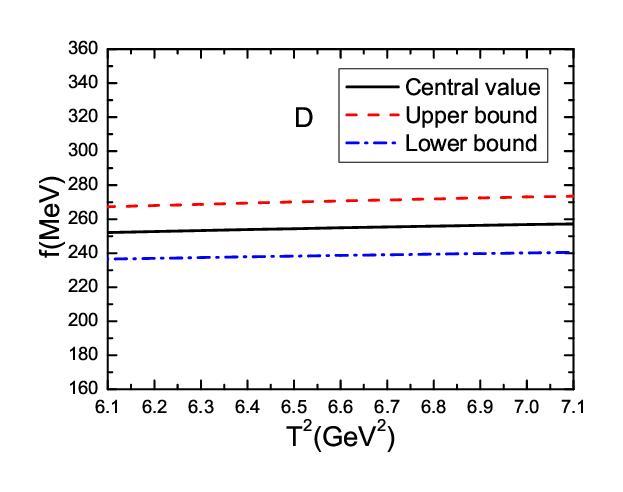}
 \includegraphics[totalheight=5cm,width=6cm]{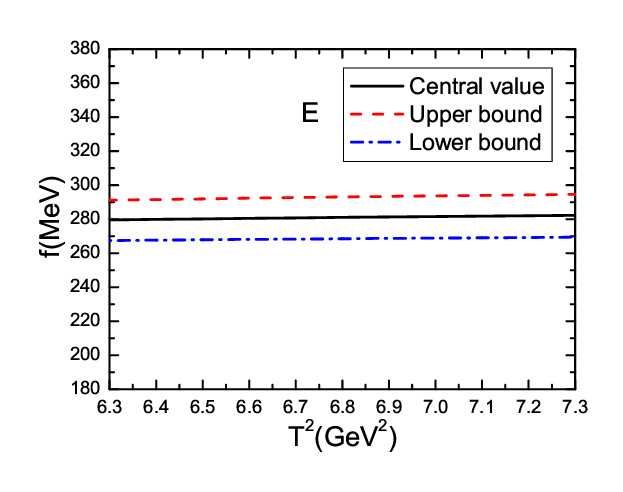}
 \includegraphics[totalheight=5cm,width=6cm]{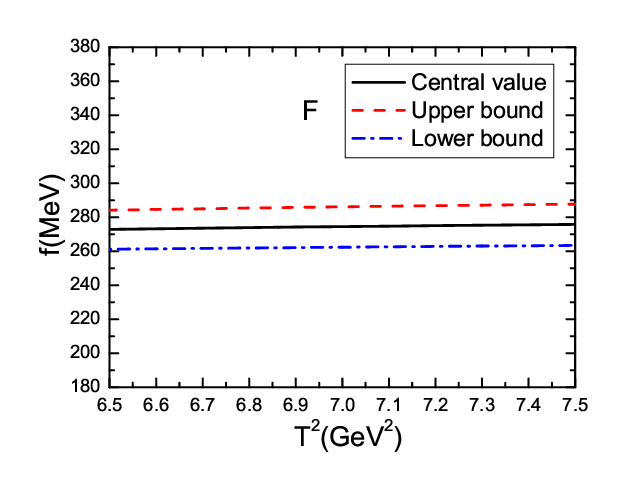}
 \includegraphics[totalheight=5cm,width=6cm]{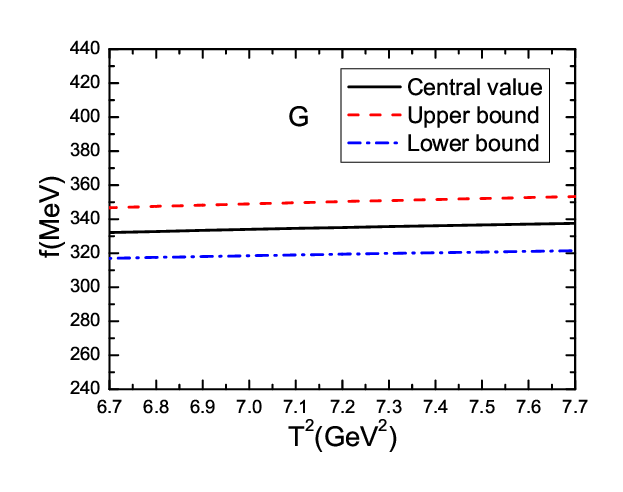}
 \includegraphics[totalheight=5cm,width=6cm]{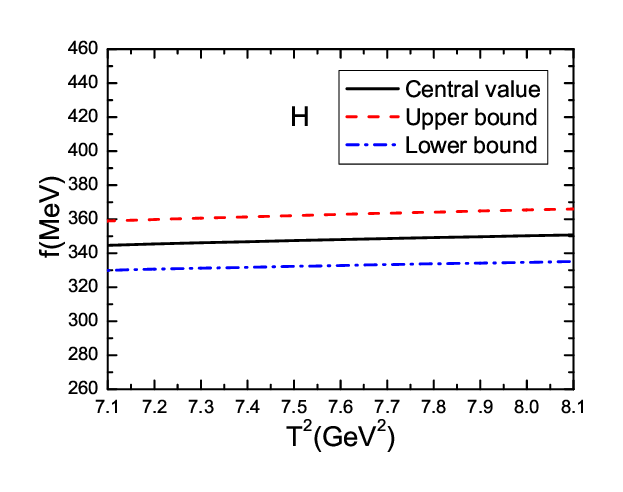}
        \caption{ The decay constants of the bottom mesons with variations of the Borel parameters $T^2$, the $A$, $B$, $C$, $D$, $E$, $F$, $G$ and $H$ denote  the mesons $B$, $B_s$, $B^*$, $B^*_s$, $B_0^*$, $B_{s0}^*$, $B_1$ and $B_{s1}$, respectively.}
\end{figure}

\begin{table}
\begin{center}
\begin{tabular}{|c|c|c|c|c|c|c|c|}\hline\hline
              & $T^2 (\rm{GeV}^2)$  & $s_0 (\rm{GeV}^2)$   & pole         & $m_{S/P/V/A}(\rm{GeV})$    & $f_{S/P/V/A}(\rm{MeV})$   \\ \hline

  $D$         & $1.2-1.8$           & $6.2\pm0.5$          & $(67-93)\%$  & $1.87\pm0.10$              & $208\pm10$          \\ \hline
 $D_{s}$      & $1.3-1.9$           & $7.3\pm0.5$          & $(77-96)\%$  & $1.97\pm0.10$              & $240\pm10$          \\ \hline

  $D^* $      & $1.9-2.5$           & $6.4\pm0.5$          & $(51-76)\%$  & $2.01\pm0.08$              & $263\pm21$          \\ \hline
 $D_{s}^*$    & $1.8-2.4$           & $7.5\pm0.5$          & $(66-87)\%$  & $2.11\pm0.07$              & $308\pm21$          \\ \hline

 $D^*_0$      & $2.4-3.0$           & $8.3\pm0.5$          & $(59-78)\%$  & $2.40\pm0.05$              & $373\pm19$          \\ \hline
 $D^*_{s0}$   & $2.1-2.7$           & $7.4\pm0.5$          & $(55-77)\%$  & $2.32\pm0.05$              & $333\pm20$          \\ \hline

 $D_1$        & $2.9-3.5$           & $8.6\pm0.5$          & $(52-70)\%$  & $2.42\pm0.05$              & $332\pm18$          \\ \hline
 $D_{s1}$     & $2.7-3.3$           & $9.3\pm0.5$          & $(61-78)\%$  & $2.46\pm0.06$              & $345\pm17$          \\ \hline

  $B$         & $5.5-6.5$           & $34.0\pm1.0$         & $(44-63)\%$  & $5.28\pm0.07$              & $194\pm15$          \\ \hline
 $B_{s}$      & $5.3-6.3$           & $36.0\pm1.0$         & $(56-74)\%$  & $5.37\pm0.07$              & $231\pm16$          \\ \hline

 $B^*$        & $6.0-7.0$           & $34.5\pm1.0$         & $(45-62)\%$  & $5.32\pm0.06$              & $213\pm18$          \\ \hline
 $B_s^*$      & $6.1-7.1$           & $36.5\pm1.0$         & $(53-69)\%$  & $5.42\pm0.06$              & $255\pm19$          \\ \hline

   $B_0^*$    & $6.3-7.3$           & $40.0\pm1.0$         & $(60-75)\%$  & $5.72\pm0.05$              & $281\pm14$          \\ \hline
 $B_{s0}^*$   & $6.5-7.5$           & $40.0\pm1.0$         & $(60-74)\%$  & $5.70\pm0.06$              & $274\pm13$          \\ \hline

 $B_1$        & $6.7-7.7$           & $41.0\pm1.0$         & $(62-75)\%$  & $5.74\pm0.05$              & $335\pm18$          \\ \hline
 $B_{s1}$     & $7.1-8.1$           & $42.0\pm1.0$         & $(63-76)\%$  & $5.76\pm0.06$              & $348\pm18$          \\ \hline
 \hline
\end{tabular}
\end{center}
\caption{ The Borel parameters, continuum threshold parameters, pole contributions, masses and decay constants of  the heavy-light  mesons. }
\end{table}

From Table 1, we can see that the experimental values of the masses of the observed mesons  $(D,D^*)$, $(D_s,D_s^*)$, $(D_0^*(2400),D_1(2430))$, $(D_{s0}^*(2317),D_{s1}(2460))$, $(B,B^*)$, $(B_s,B_s^*)$ can be well reproduced.
The masses of the  $(B^*_{0}, B_{1})$ and  $(B^*_{s0}, B_{s1})$ vary in  rather large ranges from different theoretical approaches, $m_{B^*_0}=(5.53-5.76)\,\rm{GeV}$, $m_{B_1}=(5.58-5.78)\,\rm{GeV}$, $m_{B_{s0}^*}=(5.63-5.83)\,\rm{GeV}$, $m_{B_{s1}}=(5.67-5.86)\,\rm{GeV}$, for a comprehensive review, one  can consult Ref.\cite{HYCheng}.
The present predictions $m_{B^*_0}=(5.72 \pm 0.05)\,\rm{GeV}$, $m_{B_1}=(5.74 \pm 0.05)\,\rm{GeV}$, $m_{B_{s0}^*}=(5.70 \pm 0.06)\,\rm{GeV}$, $m_{B_{s1}}=(5.76 \pm 0.06)\,\rm{GeV}$ are compatible with those values.

The thresholds of the  $DK$, $D^*K$, $BK$ and $B^*K$ states are $m_{DK}=2.36\,\rm{GeV}$, $m_{D^*K}=2.50\,\rm{GeV}$, $m_{BK}=5.78\,\rm{GeV}$ and $m_{B^*K}=5.82\,\rm{GeV}$, respectively. The $D_{s0}^*(2317)$ and $D_{s1}(2460)$ lie below the thresholds $m_{DK}$ and $m_{D^*K}$, respectively, the Okubo-Zweig-Iizuka allowed strong decays $D_{s0}^*(2317)\to DK$ and $D_{s1}(2460)\to D^*K$ are
kinematically forbidden, the widths of the $D_{s0}^*(2317)$ and $D_{s1}(2460)$ are very narrow.
According to the present predictions $m_{B_{s0}^*}=(5.70 \pm 0.06)\,\rm{GeV}$ and $m_{B_{s1}}=(5.76 \pm 0.06)\,\rm{GeV}$, the $B_{s0}^*$ and $B_{s1} $ also lie below the corresponding $BK$ and $B^*K$ thresholds,  respectively.  The strong decays $B^*_{s0}\rightarrow
BK$ and $B_{s1}\rightarrow B^*K$ are kinematically  forbidden, the
P-wave heavy mesons  $B_{s0}^*$ and $B_{s1}$ can decay through the
isospin violation precesses $B_{s0}^*\rightarrow B_s\eta\rightarrow
B_s\pi^0$ and $B_{s1}\rightarrow B_s^*\eta\rightarrow
 B_s^*\pi^0$ respectively or through the radiative decays \cite{WangBs0}. The   $\eta$ and $\pi^0$ transition matrix is very small according to
   Dashen's
 theorem \cite{Dashen},
 $ t_{\eta\pi}=\langle \pi^0 |\mathcal {H}
 |\eta\rangle=-0.003\,\rm{GeV}^2$,
the P-wave bottomed mesons $B_{s0}^*$ and $B_{s1}$, just like their
charmed cousins $D_{s0}^*(2317)$ and $D_{s1}(2460)$, maybe very narrow
\cite{WangDs0}. The present predictions are consistent with our previous work \cite{WangCPL}, but the analysis is refined by including more terms in the operator product expansion.

The values of the decay constants of the pseudoscalar mesons are slightly different from the ones  in  our previous work \cite{WangJHEP}.
In Table 2, we compare the present predictions to the experimental data and other theoretical calculations, such the QCD sum rules (QCDSR) \cite{QCDSRfD2,QCDSRfD5,QCDSRfD4,QCDSRfD6,SRBaker} and lattice QCD (LQCD) \cite{LattfD1,LattfD2,LattfD3}. The present  predictions $f_D=(208 \pm 10)\,\rm{MeV}$ and $f_B=(194 \pm 15)\,\rm{MeV}$ are consistent with the experimental data within uncertainties, while the prediction  $f_{D_s}=(240 \pm 10)\,\rm{MeV}$ is lies below the lower bound of the experimental value  $f_{D_s}=(257.5 \pm 4.6)\,\rm{MeV}$ \cite{PDG}.
We take the $\overline{MS}$ mass $m_c(\mu)$ and truncate the perturbative corrections to the order $\mathcal{O}(\alpha_s)$, the experimental values of the $f_{D}$, $f_{D_s}$ and $f_{D_s}/f_{D}$ cannot be reproduced consistently by  the QCD sum rules.
The existence of a charged Higgs boson or any other charged
object beyond the standard model would modify the decay rates, see Eq.(1), therefore   modify the values of the decay constants, for example, the leptonic decay widths are modified in two-Higgs-doublet models \cite{Higgs}. If the predictions of the $f_{D}$, $f_{D_s}$ and $f_{D_s}/f_{D}$ based on the QCD sum rules   are  close to the true values, new physics beyond the standard model are favored so as to smear the discrepancies  between the theoretical calculations and experimental data.

The analytical expression of the  perturbative  $\mathcal{O}(\alpha_s)$ corrections $R_5\left(\frac{m_Q^2}{s}\right)$ is well known \cite{Aliev-pert}, while the
semi-analytical perturbative  $\mathcal{O}(\alpha_s^2)$ corrections are presented as mathematical functions ${\rm R2sFF} [v]$, ${\rm R2sFA} [v]$, ${\rm R2sFL} [v]$ and
$ {\rm R2sFH}[v]$ with the variable $v=\left(1-\frac{m_Q^2}{s} \right)/\left(1+\frac{m_Q^2}{s} \right)$ at the energy-scale of the heavy quark pole mass $\mu=m_Q$ \cite{QCDSR-3loop}. The analytical expressions of the  terms which contain  logarithms  such as $\log\frac{\mu^2}{m_Q^2}$, $\log\frac{\mu^2}{s}$ cannot be recovered,
   it is unreasonable to take other energy scale besides $m_Q$. Now we choose the pole masses $m_Q$ and take into account  the semi-analytical perturbative  $\mathcal{O}(\alpha_s^2)$ corrections by  setting  $n_f=4$ and $\mu=m_c$ for the $D$ ($D_s$) meson and $n_f=5$ and $\mu=m_b$ for the $B$ ($B_s$) meson.

The on-shell quark propagators have no infrared divergences in perturbation theory, which
  provides a perturbative definition of the quark masses.  The full quark
propagators have no poles  because the quarks are confined, so   the pole masses cannot be defined outside of perturbation theory.
 Furthermore, the pole masses cannot be used to arbitrarily high accuracy because  nonperturbative infrared effects in QCD. We choose the pole masses just because the  semi-analytical perturbative  $\mathcal{O}(\alpha_s^2)$ corrections are calculated by taking  the pole mass $m_Q$ and setting the energy scale  to be  $\mu=m_Q$ \cite{QCDSR-3loop}.
   The contributions of the  $u$, $d$ masses are tiny and can be neglected safely.
  In calculations, we   set the pole masses $m_u=m_d=0$, $m_s=150\,\rm{MeV}$,  and observe that
the  masses of the heavy pseudoscalar mesons increase  monotonously with increase  of the pole masses, the  values of the pole masses  $m_c=1.44\,\rm{GeV}$ and $m_b=4.67\,\rm{GeV}$ can lead to satisfactory values by choosing reasonable Borel parameters and threshold parameters. Those pole masses are different from the $\overline{MS}$ masses, for example, $m_c(\mu=1\,\rm{GeV})=1.39\,\rm{GeV}$, $m_b(\mu=1\,\rm{GeV})=6.07\,\rm{GeV}$, $m_c(\mu=2\,\rm{GeV})=1.13\,\rm{GeV}$, $m_b(\mu=2\,\rm{GeV})=4.87\,\rm{GeV}$ from Eq.(29). The pole masses are energy scale independent, therefore the energy scale dependence of the QCD spectral densities originate only  from the vacuum condensates.

The  Borel parameters, continuum threshold parameters, pole contributions, and the resulting masses and decay constants of the heavy pseudoscalar mesons are shown in Table 3, the values are slightly different from the ones in our previous work \cite{WangJHEP}.
From Table 1 and Table 3, we can see that the present  predictions $f_D=(210 \pm 11)\,\rm{MeV}$, $f_{D_s}=(259 \pm 10)\,\rm{MeV}$ and $f_B=(192 \pm 13)\,\rm{MeV}$ are  in excellent agreement with  the experimental data within uncertainties \cite{PDG}. The ratio $f_{D_s}/f_D=1.23\pm0.07$ is also in excellent agreement with  the experimental data $f_{D_s}/f_D=1.258 \pm 0.038$ \cite{PDG}, which indicates that  the perturbative $\mathcal{O}(\alpha_s^2)$ corrections should be taken into account. However, the pole masses $m_Q$ and energy scales $\mu=m_Q$ have be chosen, as the semi-analytical expressions are obtained  at such conditions.   In this case,  new physics beyond the standard model are not favored, as the agreements between the experimental data and present theoretical calculations are already excellent.

In Table 4, we compare the present predictions for the decay constants of the heavy vector mesons to other theoretical  calculations, such as the QCD sum rules \cite{QCDSRfD6,QCDSRfDv1,Narison1404}, lattice QCD \cite{LattfD-Dv1,LattfDv1,LattfDv2,LattfDv3}, the relativistic potential model (RPM) \cite{RPMfD1},
the field-correlator method (FCM) \cite{FCMfD1}, and the light-front quark model \cite{LFQMfDv1}. From the table, we can see that the predictions differ from each other in one  way or  the  other.
In Table 5, we compare the present predictions for the decay constants of the heavy scalar mesons to the ones from the QCD sum rules \cite{QCDSRfD01} and lattice QCD \cite{LattfD01}. From the table, we can see that the predictions are consistent with the ones from lattice calculations but  differ greatly from the ones from the QCD sum rules.

If we turn off the perturbative $\mathcal{O}(\alpha_s)$ corrections to the quark condensates and choose the same parameters, such as the $\overline{MS}$ masses, Borel parameters and continuum threshold parameters, etc, the masses and decay constants undergo reduction or increment in a definite way according to the spin and parity, see Table 6. From the table, we can see that the mass-shifts of the $D$-mesons with $J^P=0^\pm$ are larger than $40\,\rm{MeV}$, while the shifts of the masses and decay constants of   all the $B$-mesons are small and can be neglected. We can    re-choose the Borel windows to warrant the mass-shifts $\delta m_{S/P/V/A}=0$,  and account for the net effects by the shifts of the decay constants $\delta f_{S/P/V/A}$, which are shown the bracket in Table 6. From the table, we can see that the largest shift of the decay constant $\delta f_{D}=-11\,\rm{MeV}$, which exceeds the total uncertainty of the decay constant $\delta f_D=\pm 10\,\rm{MeV}$ (see Table 1), the shifts of the decay constants of the $D$-mesons with $J^P=0^\pm,\,1^-$ are larger than $5\,\rm{MeV}$, while for other mesons, the shifts of the decay constants $|\delta f|\leq 4\,\rm{MeV}$.
All in all, we should take into account the perturbative $\mathcal{O}(\alpha_s)$ corrections to the quark condensates in a comprehensive study.

\begin{table}
\begin{center}
\begin{tabular}{|c|c|c|c|c|c|c|}\hline\hline
                      &$f_{D}(\rm{MeV})$     &$f_{D_{s}}(\rm{MeV})$ &$f_{B}(\rm{MeV})$     &$f_{B_{s}}(\rm{MeV})$ &$f_{D_{s}}/f_{D}$     &$f_{B_{s}}/f_{B}$\\ \hline
 Expt \cite{PDG}      &$204.6\pm 5.0$        &$257.5 \pm 4.6$       &$190.6\pm4.7$         &                      &$1.258\pm 0.038$      &                \\ \hline
 QCDSR \cite{QCDSRfD2}&$177\pm 21$           &$205\pm22$            &$178 \pm 14 $         &$200\pm14$            &$1.16\pm0.16$         &$1.12\pm0.11$        \\ \hline

 QCDSR \cite{QCDSRfD5}&$204\pm6$             &$246\pm6$             &$207\pm8$             &$234\pm5$             &$1.21\pm0.04$         &$1.14\pm0.03$   \\ \hline
 QCDSR \cite{QCDSRfD4}&$206.2\pm7.3$         &$245.3\pm15.7$        &$193.4\pm12.3$        &$232.5\pm18.6$        &$1.193\pm0.025$       &$1.203\pm0.020$   \\ \hline

 QCDSR \cite{QCDSRfD6}&$201^{+12}_{-13}$     &$238^{+13}_{-23}$     &$207^{+17}_{-09}$     &$242^{+17}_{-12}$     &$1.18^{+0.04}_{-0.05}$&$1.17^{+0.03}_{-0.04}$\\ \hline
 QCDSR \cite{SRBaker} &                      &                      &$186\pm14$            &$222\pm12$            &                      &$1.19\pm0.09$\\ \hline

 LQCD \cite{LattfD1}  &$197\pm9$             &$244\pm8$             &                      &                      &$1.24\pm0.03  $       &                 \\ \hline
 LQCD \cite{LattfD2}  &$213\pm4$             &$248.0\pm2.5$         &$191\pm9$             &$228\pm10$            &$1.164\pm0.018$       &$1.188\pm0.018$ \\ \hline
 LQCD \cite{LattfD3}  &$218.9\pm11.3$        &$260.1\pm10.8$        &$196.9\pm8.9$         &$242.0\pm9.5$         &$1.188\pm0.025$       &$1.229\pm0.026$ \\ \hline
 This work            &$208 \pm 10$          &$240 \pm 10$          &$194 \pm 15$          &$231 \pm 16$          &$1.15\pm0.06$         &$1.19\pm0.10$   \\ \hline
 This work${}^*$      &$210\pm11$            &$259\pm10$            &$192\pm13$            &$230\pm13$            &$1.23\pm0.07$         &$1.20\pm0.09$   \\ \hline
 \hline
\end{tabular}
\end{center}
\caption{ The decay constants of the heavy pseudoscalar mesons from the  experimental data, the QCD sum rules and lattice QCD, the superscript star $*$ denotes that the pole masses are chosen  and perturbative $\mathcal{O}(\alpha_s^2)$ corrections are taken into account.}
\end{table}

\begin{table}
\begin{center}
\begin{tabular}{|c|c|c|c|c|c|c|c|}\hline\hline
              & $T^2 (\rm{GeV}^2)$  & $s_0 (\rm{GeV}^2)$   & pole         & $m_P(\rm{GeV})$    & $f_{P}(\rm{MeV})$   \\ \hline
   $D $       & $1.4-2.0$           & $5.5\pm0.5$          & $(55-85)\%$  & $1.87\pm0.06$      & $210\pm11$          \\ \hline
 $D_{s}$      & $1.0-1.6$           & $7.4\pm0.5$          & $(86-98)\%$  & $1.97\pm0.07$      & $259\pm10$          \\ \hline
   $B$        & $4.1-4.9$           & $33.0\pm1.0$         & $(55-75)\%$  & $5.28\pm0.04$      & $192\pm13$          \\ \hline
 $B_{s}$      & $4.4-5.2$           & $35.0\pm1.0$         & $(61-79)\%$  & $5.37\pm0.04$      & $230\pm13$          \\ \hline
 \hline
\end{tabular}
\end{center}
\caption{ The Borel parameters, continuum threshold parameters, pole contributions, masses and decay constants of the heavy pseudoscalar mesons when the perturbative $\mathcal{O}(\alpha_s^2)$ corrections are taken into account. }
\end{table}

\begin{table}
\begin{center}
\begin{tabular}{|c|c|c|c|c|c|c|}\hline\hline
                          &$f_{D^*}(\rm{MeV})$   &$f_{D_{s}^*}(\rm{MeV})$ &$f_{B^*}(\rm{MeV})$   &$f_{B_{s}^*}(\rm{MeV})$  \\ \hline
 QCDSR \cite{QCDSRfD6}    &$242^{+20}_{-12}$     &$293^{+19}_{-14}$       &$210^{+10}_{-12}$     &$251^{+13}_{-16}$       \\ \hline
 QCDSR \cite{QCDSRfDv1}   &$252.2\pm 22.3\pm 4$  &$305.5\pm 26.8\pm 5$    &$181.8\pm13.1\pm4$    &$213.6\pm18.2\pm6$       \\ \hline
 QCDSR \cite{Narison1404} &$250\pm 11$           &$270\pm19$              &$209\pm8$             &$220\pm9$         \\ \hline
 LQCD \cite{LattfD-Dv1}   &$278 \pm 13 \pm10$    &$311 \pm 9$             &                      &                        \\ \hline
 LQCD \cite{LattfDv1}     &                      &$274 \pm 6$             &                      &                        \\ \hline
 LQCD \cite{LattfDv2}     &                      &                        &$175\pm6$             &$213\pm7$               \\ \hline
 LQCD \cite{LattfDv3}     &$245\pm20$            &$272\pm16$              &$196\pm24$            &$229\pm20$               \\ \hline

 RPM  \cite{RPMfD1}       &$310$                 &$315$                   &$219$                 &$251$                 \\ \hline
 FCM \cite{FCMfD1}        &$273\pm13$            &$307\pm18$              &$200\pm10$            &$230\pm12$                 \\ \hline
 LFQM \cite{LFQMfDv1}     &$245^{+35}_{-34}$     &$272^{+39}_{-38}$       &$196^{+28}_{-27}$     &$229^{+32}_{-31}$                 \\ \hline

 This work                &$263 \pm 21$          &$308 \pm 21$            &$213 \pm 18$          &$255 \pm 19$           \\ \hline
 \hline
\end{tabular}
\end{center}
\caption{ The decay constants of the heavy vector mesons from the   some theoretical calculations.}
\end{table}

\begin{table}
\begin{center}
\begin{tabular}{|c|c|c|c|c|c|c|}\hline\hline
                         &$f_{D^*_0}(\rm{MeV})$ &$f_{D_{s0}^*}(\rm{MeV})$ &$f_{B^*_0}(\rm{MeV})$ &$f_{B_{s0}^*}(\rm{MeV})$  \\ \hline
 QCDSR \cite{QCDSRfD01}  &                      &$128 \pm 13$             &                      &                        \\ \hline
 LQCD \cite{LattfD01}    &$360\pm 90$           &$340\pm 110$             &                      &                        \\ \hline

 This work               &$373\pm 19$           &$333 \pm 20$             &$281 \pm 14$          &$274 \pm 13$           \\ \hline
 \hline
\end{tabular}
\end{center}
\caption{ The decay constants of the heavy scalar mesons from the  some theoretical calculations.}
\end{table}

\begin{table}
\begin{center}
\begin{tabular}{|c|c|c|c|c|c|c|c|}\hline\hline
                      & $\delta m_{S/P/V/A}(\rm{MeV})$    & $\delta f_{S/P/V/A}(\rm{MeV})$   \\ \hline

  $D$    ($0^-$)      & $+42$                             & $-9$  ($-11$)        \\ \hline
 $D_{s}$ ($0^-$)      & $+34$                             & $-6$  ($-8$)        \\ \hline

  $D^* $ ($1^-$)      & $+22$                             & $-6$ ($-8$)         \\ \hline
 $D_{s}^*$ ($1^-$)    & $+12$                             & $-3$ ($-5$)         \\ \hline

 $D^*_0$  ($0^+$)     & $-43$                             & $+2$ ($+6$)         \\ \hline
 $D^*_{s0}$ ($0^+$)   & $-44$                             & $+1$ ($+6$)        \\ \hline

 $D_1$    ($1^+$)     & $-5$                              & $+1$ ($+3$)         \\ \hline
 $D_{s1}$  ($1^+$)    & $-2$                              & $+1$ ($+2$)         \\ \hline

  $B$    ($0^-$)      & $+2$                              & $-3$ ($-4$)          \\ \hline
 $B_{s}$ ($0^-$)      & $+1$                              & $-2$ ($-3$)             \\ \hline

 $B^*$  ($1^-$)       & $+0$                              & $-3$ ($-3$)          \\ \hline
 $B_s^*$ ($1^-$)      & $+0$                              & $-2$ ($-2$)          \\ \hline

   $B_0^*$  ($0^+$)   & $-2$                              & $+1$ ($+3$)          \\ \hline
 $B_{s0}^*$  ($0^+$)  & $-1$                              & $+1$ ($+3$)          \\ \hline

 $B_1$       ($1^+$)  & $-0$                              & $+3$ ($+3$)          \\ \hline
 $B_{s1}$    ($1^+$)  & $-0$                              & $+2$ ($+2$)          \\ \hline
 \hline
\end{tabular}
\end{center}
\caption{ The shifts of the masses and decay constants of  the heavy-light  mesons when the perturbative $ \mathcal{O} (\alpha_s)$  corrections to the quark condensates are turned  off. We can  re-choose the Borel windows to warrant the mass-shifts  $\delta m_{S/P/V/A}=0$, the resulting shifts of the decay constants are shown in the bracket. The $+0$ ($-0$) denotes the value $0<\delta m <1\,\rm{MeV}$ ($-1\,{\rm{MeV}}<\delta m <0$).  }
\end{table}

\section{Conclusion}
In this article, we calculate the contributions of the vacuum condensates up to dimension-6, in including  the $\mathcal{O}(\alpha_s)$ corrections  to the quark condensates,  in the operator product expansion. Then we study the masses and decay constants of the  pseudoscalar, scalar, vector and axial-vector heavy-light mesons with the QCD sum rules in a systematic way. In calculations, we take the $\overline{MS}$ masses and take into account the perturbative $\mathcal{O}(\alpha_s)$ corrections. The masses of the observed heavy-light mesons  $(D,D^*)$, $(D_s,D_s^*)$, $(D_0^*(2400),D_1(2430))$, $(D_{s0}^*(2317),D_{s1}(2460))$, $(B,B^*)$, $(B_s,B_s^*)$ can be well reproduced, while the predictions for the masses of
 the  $(B^*_{0}, B_{1})$ and  $(B^*_{s0}, B_{s1})$ can be confronted with the experimental data in the futures. Up to the order $\mathcal{O}(\alpha_s)$, the QCD sum rules cannot lead to satisfactory values for the $f_{D}$, $f_{D_s}$ and $f_{D_s}/f_{D}$ compared to the experimental data. We have to take into account the perturbative $\mathcal{O}(\alpha_s^2)$ corrections by choosing the pole masses, then the experimental data can be well reproduced. The present predictions for the decay constants of the  heavy-light pseudoscalar, scalar, vector and axial-vector mesons have many phenomenological applications in studying the semi-leptonic and leptonic decays  of the
 heavy-light mesons.

\section*{Acknowledgements}
This  work is supported by National Natural Science Foundation,
Grant Numbers 11375063, and Natural Science Foundation of Hebei province, Grant Number A2014502017.

\end{document}